\documentclass[11pt,oneside,onecolumn]{article}
\usepackage[]{fontenc}
\usepackage[dvips]{epsfig}
\usepackage{a4}

\makeatletter
\def\LyX{L\kern-.1667em\lower.25em\hbox{Y}\kern-.125emX\spacefactor1000}%
\newcommand{\lyxtitle}[1] {\thispagestyle{empty}
\global\@topnum\z@
\section*{\LARGE \centering \sffamily \bfseries \protect#1 }
}

\newenvironment{lyxbibliography}
{
\begin{thebibliography}{99}}
{\end{thebibliography}}

\makeatother

\begin{document}

\title{The state space of short-range Ising spin glasses: the density of
states}

\author{T. Klotz, S. Schubert, K. H. Hoffmann\\
Inst. f. Physik, TU Chemnitz, D-09107 Chemnitz, Germany}

\maketitle

\begin{abstract}

The state space of finite square and cubic Ising spin glass models
is analysed in terms of the global and the local density of states.
Systems with uniform and gaussian probability distribution of interactions
are compared. Different measures for the local state density are presented
and discussed. In particular the question whether the local density
of states grows exponentially or not is considered. The direct comparison
of global and local densities leads to consequences for the structure
of the state space.

\end{abstract}

\section{Introduction}

The often very unusual dynamic behaviour of complex systems like spin
glasses \cite{Aging1,Aging2} is significantly determined by the properties
of their state space. One key to understand the relaxation and aging
effects in this class of systems in particular for the low-temperature
region is given by the structure of local minima and barriers in the
low-lying energy landscape. In order to construct models of this landscape,
which are useful for simulating the non-equilibrium dynamics, it is
necessary to extract and to quantify the important structural properties
of this landscape. Unfortunately in experiments the state-space structure
is only indirectly accessable. For systems with long-range interactions
analytical mean-field-like methods have been applied to investigate
the state-space properties. The situation for short-range systems
is more complicated. Due to the computational effort needed the exact
calculation of the energetically low-lying excitations of a system
is restricted to small system sizes. Nevertheless the analysis of
small sytems can give a first understanding of effects in principle
and help to check model assumptions.

One of the simplest models for complex systems is the Ising-spin glass.
There has been done a lot of research concerning the long-range SK-model
\cite{SK}. A few numerical and experimental works tried to analyse
the state-space structure more or less directly \cite{Nemoto,Vertechi}.
This was mostly done in order to check the interesting theoretical
predictions for the hierarchical structure of the phase space of the
SK model \cite{Ultrametricity}. For short-range systems the situation
is more unsatisfying. For small \( \pm J \) model systems a detailed analysis
of the state-space has be done in \cite{Klotz}. It is unclear how
strong the state-space structure found is influenced by the discretness
of the interactions. As a counterpart to the \( \pm J \) systems usually systems
with gaussian distributed interactions between nearest neighbours
are treated. An extensive analysis of the morphology of the state
space was undertaken in \cite{Sibani} by use of the so called lid
method. 

An interesting outcome of this investigations was, that the local
density of states inside a state-space pocket seems to grow exponentially.
Such an exponential increase of the local density of states with increasing
energy would lead to drastical thermodynamic consequences. For temperatures
below a critical temperature the occupation probability would reach
its maximum at the ground-state energy. Thus the system is trapped
in a certain state-space valley for a long time or even forever, provided
that the barriers surrounding this pocket are high enough. For increasing
temperatures the maximum of the occupation probability jumps at a
critical temperature \( T_{\mathrm{c}} \) from the minimal to the maximal energy of the
system. Therefore the probability to leave the considered valley increases
drastically. The system is no longer trapped in this valley. 

This behaviour is not only important from a thermodynamic point of
view, but for optimization problems and methods, too \cite{sven}.
Assuming there exists such a critical temperature, the cooling scheme
for simulated annealing methods should be chosen in such a way, that
the algorithm has found the ground-state valley at a temperature above
the critical one. Otherwise it may happen, that the algorithm never
finds the true ground-state due to the low probability to jump to
other valleys below the critical temperature.

If the local density of states is exactly exponential, the critical
temperature is sharply defined. However, if there is no exponential
behaviour, the transition might vanish or is at least smeared out.
In this paper we try to clearify this situation. We analysed finite
two- and three-dimensional systems with respect to their density of
states. Starting from the exact knowledge of all energetically low-lying
states we calculated at first the global density of states. After
sorting the states according to the valley in state space they belong
to, we will discuss various differently measures for the local state
density. Finally we will compare these different measures.

\section{Model and methods}

In the following we will present results for two- and three-dimensional
Ising-spin systems on square and cubic lattices with randomly chosen
interactions between nearest neighbours and periodic boundary conditions.
The lattice size is restricted by computational reasons and is \( L=8 \) for
the two-dimensional and \( L=4 \) for the three-dimensional case. There is
no external field applied to the systems.

We analysed systems with a gaussian distribution of interactions as
well as systems with a uniform distribution. A disantvantage of the
gaussian distribution of interactions in particular for local structure
investigations of the state space is the possibility of extremely
large local fields. These fields can lead to a crossing of all energy
barriers by just a single spin flip. This unphysical drawback can
be overcome by using a uniform distribution, which is in this sense
a counterpart to the gaussian one. It restricts the maximal strengths
of interactions and thus the maximal local field. 

In order to allow the comparison of both distributions the first two
moments have been set equal. As usual the mean is set to zero and
the standard deviation is normalized to unity. If this choice leads
to a very similar state-space structure, both system classes could
be used alternatively.

The basis of the state-space analysis is an exact determination of
all energetically low-lying states up to a given cut-off energy by
the method of recursive branch-and-bound \cite{BB}. The main idea
of this method is to search the binary tree of all states. The search
can be restricted by finding lower bounds for the minimal reachable
energy inside of a subtree. If this lower bound is higher than the
energy of a suboptimal state already found, it is not necessary to
examine the corresponding subtree. A first good suboptimal state can
be found by recursively solving smaller subproblems. By adding an
energy offset to the calculated lower bounds it is possible to calculate
not only the ground states, but all states below a given cut-off energy
too.

The obtained states were ordered by increasing energy using a distributed
sort algorithm. Starting from the ground state and successively increasing
the maximal energy of the considered states all states are sorted
according to their valley in the configuration space. For a chosen
energy two states are sorted to the same valley, if one state can
be reached from the other via a series of single spin flips without
exceeding the chosen energy. Thus the definition of a valley depends
on this energy. Each valley can be addressed by the state with minimal
energy inside of the valley. Note, that a valley is joined with a
more low-lying valley, if the considered energy becomes larger than
the barrier between both valleys. Furthermore it should be noted here,
that the definition of a valley of course depends on the definition
of neighbouring spin configurations. As it is done in most investigations
we restricted ourself to consider only single spin flip processes.

\section{Results}

\begin{figure}
{\centering \epsfig{file=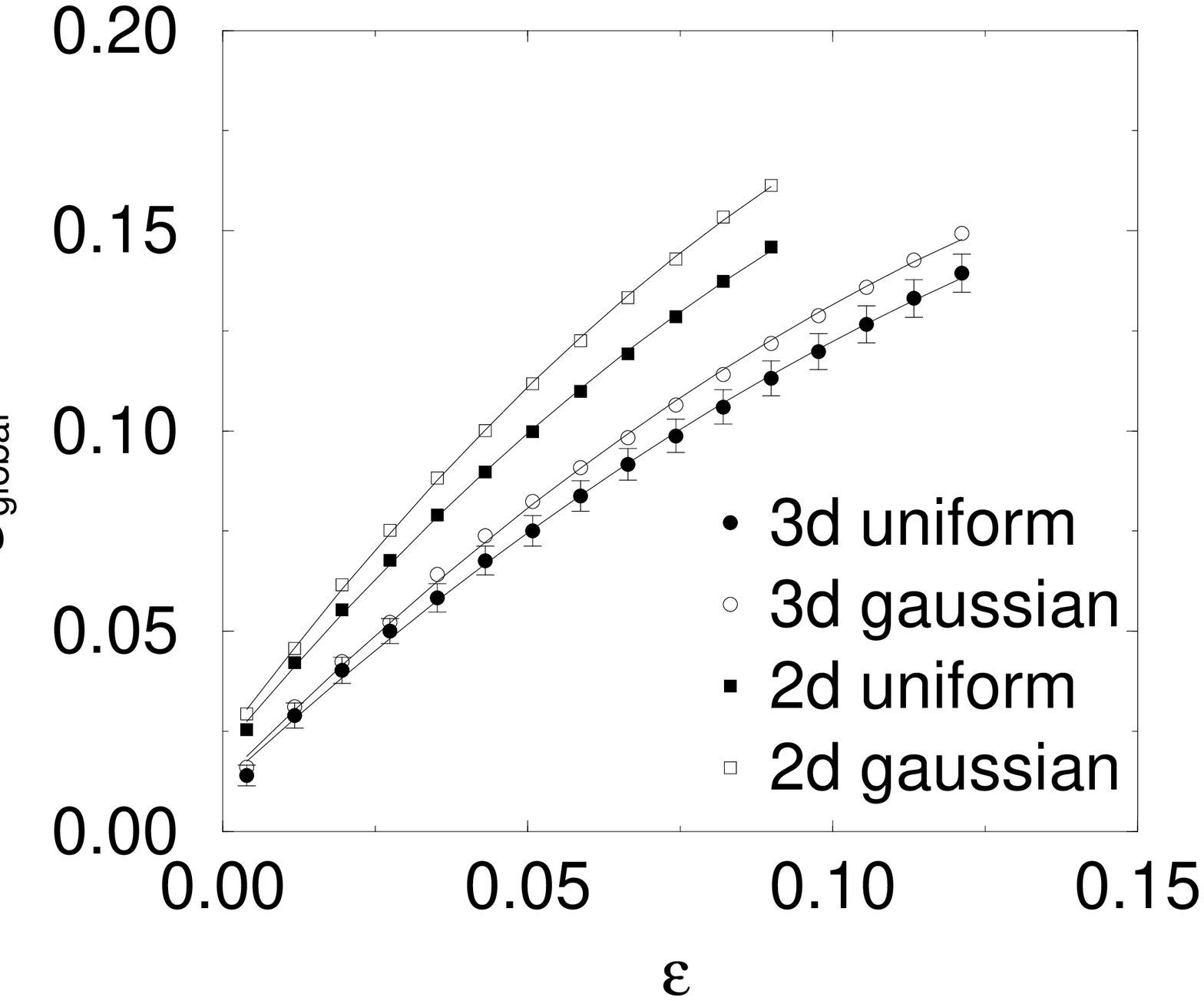, width=1\textwidth} \par}

\caption{Global density of states \protect\( g_{\mathrm{global}}\left( \varepsilon \right) \protect \) for \protect\( 2d\protect \) (squares) and \protect\( 3d\protect \) (circles) systems
with uniform (full symbols) or gaussian (open symbols) distribution
of interactions averaged over different realizations of interactions.
The lines correspond to quadratic fits.\label{fig:g_global}}
\end{figure}

The global density of states \( g_{\mathrm{global}}\left( \varepsilon \right)  \) (GDOS) is defined as the number of
states with energy \( \varepsilon  \) per spin above the ground state. Fig. \ref{fig:g_global} shows
the logarithm of the global density of states normalized to the number
of spins \( N \) in the system for the \( 2d \) and the \( 3d \) systems with gaussian
and uniform distribution. The results are averaged over 20 different
realizations of disorder for the gaussian distribution and over 50
samples for the uniform distribution. The errorbars give an idea of
the sample to sample fluctuations.

\begin{figure}
{\centering \epsfig{file=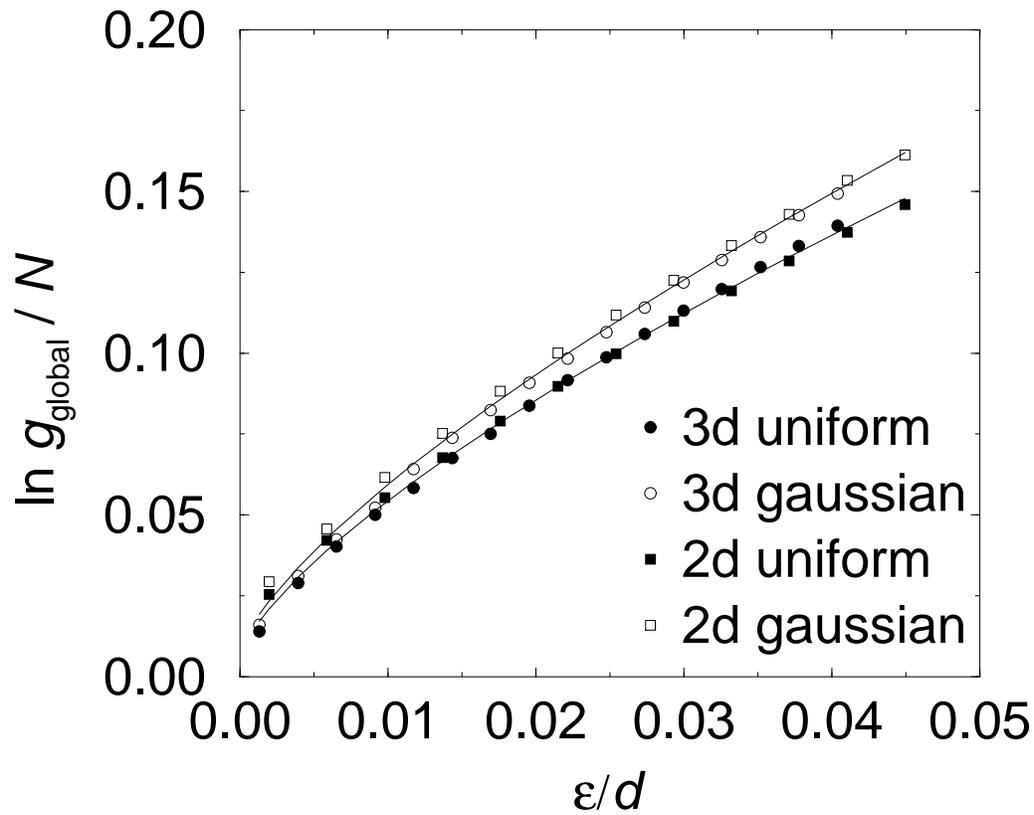, width=1\textwidth} \par}

\caption{Global density of states \protect\( g_{\mathrm{global}}\protect \) versus energy per bond for \protect\( 2d\protect \) (squares)
and \protect\( 3d\protect \) (circles) systems with uniform (full symbols) or gaussian (open
symbols) distribution of interactions averaged over different realizations
of interactions. The lines correspond to power fits.\label{fig:g_global_bond}}
\end{figure}

The GDOS is significantly higher for the \( 2d \) systems compared to the
\( 3d \) systems. This is obviously caused by the different coordination
numbers, as can be seen in fig. \ref{fig:g_global_bond}. The GDOS for the gaussian distribution
is slightly higher than for the uniform distribution. Nevertheless,
there seems to be no qualitative difference between both curves. 

For all systems the GDOS increases clearly subexponentially with energy.
To quantify this behaviour it is possible to make an ansatz of the
form 
\begin{equation}
\label{eq:g_ansatz}
g\left( \varepsilon \right) \propto \exp \left( c+\alpha \varepsilon +\gamma \varepsilon ^{\delta }\right) 
\end{equation}
for small energies \( \varepsilon  \) above the minimal energy. The occupation
probability in equilibrium then reads 
\begin{equation}
\label{eq:occupation}
p\left( \varepsilon \right) \propto \exp \left[ c+\left( \alpha -\beta \right) \varepsilon +\gamma \varepsilon ^{\delta }\right] ,
\end{equation}
where \( \beta  \) denotes the inverse
temperature. The extremal value of such a distribution is reached
for 
\begin{equation}
\label{eq:e_ext}
\varepsilon _{\mathrm{ext}}=\left( \frac{\beta -\alpha }{\gamma \delta }\right) ^{1/\left( \delta -1\right) }.
\end{equation}
The only singular point in (\ref{eq:e_ext}) is at \( \beta =\alpha  \). In the linear case \( \gamma =0 \)
the maximum of (\ref{eq:occupation}) jumps at this value of \( \beta  \) from the maximal energy
of the system for high temperatures to the minimal energy for low
temperatures. 

If however \( \gamma \neq 0 \) and \( \delta >1 \) the subexponential behaviour of the DOS as found
in our data leads to a negative coefficient \( \gamma  \). Then it can easily
be seen that the energy of the maximum of (\ref{eq:occupation}) is positive and finite
for high temperatures and goes down with decreasing temperature. At
and below \( T=1/\alpha  \) the occupation probability is highest for \( \varepsilon _{\mathrm{ext}}=0 \). If the linear
term in \( g\left( \varepsilon \right)  \) vanishes (\( \alpha =0 \)) and \( 0<\delta <1 \), the maximum energy goes continuously
from the maximal energy of the system down to the minimal one with
decreasing temperature. Thus there is no sign of a critical behaviour
caused by the DOS.

As a result of the above discussion we chose two different ansatzes
for fitting functions in order to analyse our numerical data. For
\( \delta =2 \) (\ref{eq:g_ansatz}) simplifies to a quadratic polynomial ansatz, which we call in
the following the quadratic fit. The choice \( \alpha =0 \) leads to a fitting ansatz
without any linear term, which will be called power fit.

The quadratic fits shown in fig. \ref{fig:g_global} fit the data quite well. The ratio
of the linear and the quadratic coefficient corresponds for the \( 3d \)
systems to an energy of about \( 0.47 \) per spin. The inverse linear coefficients
correspond to a temperature of \( T\sim 0.5 \) in \( 2d \) and \( T\sim 0.65 \) in \( 3d \). The errors of these
fit parameters have been estimated. For the energy per spin it is
of the order \( 0.05 \) per spin, the error for the temperature can be estimated
to \( 0.05 \). The lines in fig. \ref{fig:g_global_bond} are power fits, which seem to fit the data
quite well, too. The gaussian and the uniform distribution differ
only in the coefficients \( c \) and \( \gamma  \). The exponents \( \delta  \) are about 0.72 for
both distributions. The error is of the order \( 0.05 \).

The local DOS (LDOS) is given by the number of states inside a valley
at a given energy. In order to average the LDOS, it is necessary to
clearify the measuring procedure. We discuss here three different
possibilities. 

The first one is to start at a high temperature and to perform a steepest
descent algorithm. The LDOS of the valley the system was trapped in,
can then be measured relatively to the minimal energy of this valley.
The averaging will be done over different runs and different realizations
of interactions. We will call the measure defined in that way relatively
measured LDOS (RLDOS) and denote it by \( g_{\mathrm{rel}} \). 

The second possibility assumes that the ground-state of the system
is known already. Then the local density of states can be measured
relatively to the ground-state energy instead of the minimal energy
of the valley found. We will call this variant absolutely measured
LDOS (ALDOS) and denote it by \( g_{\mathrm{abs}} \). If the averaging procedures for \( g_{\mathrm{rel}} \)
and \( g_{\mathrm{abs}} \) are restricted to the ground-state valley, both variants are
equivalent and result in the averaged local density of ground-state
valleys (GLDOS), which will be denoted by \( g_{\mathrm{gs}} \).

It should be noted here, that in practice the averaging will be performed
not over different runs, but over all valleys found up to the cut-off
energy. This may cause an systematic error due to valleys with local
minima higher than the cut-off energy. Obviously, this effect could
only be important for \( g_{\mathrm{rel}} \).

\begin{figure}
\epsfig{file=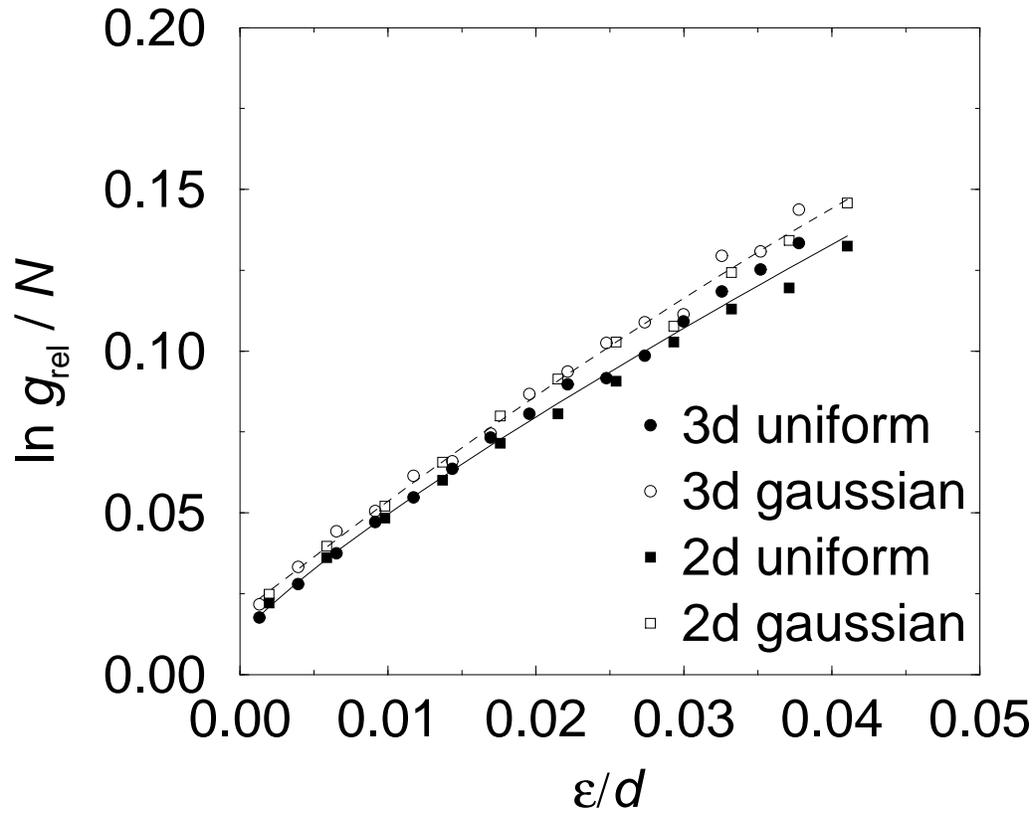, width=1\textwidth}

\caption{Relatively measured LDOS \protect\( g_{\mathrm{rel}}\protect \) versus energy per bond for \protect\( 2d\protect \) (squares)
and \protect\( 3d\protect \) (circles) systems with uniform (full symbols) or gaussian (open
symbols) distribution of interactions averaged over different configuration
space valleys and different realizations of interactions. The dashed
and the full line correspond to a quadratic fit and a power fit, respectively.\label{fig:g_local_bond_fitted}}
\end{figure}

As for the GDOS the RLDOS as a function of the energy per bond \( \varepsilon /d \) is
quite equivalent for \( 2d \) and \( 3d \) systems (Fig. \ref{fig:g_local_bond_fitted}). However, for energies
higher than \( 0.02 \) per bond there seem to be systematic deviations. In
both dimensions \( g_{\mathrm{rel}} \) is slightly higher for the gaussian distribution.
Both fitting ansatzes fit the numerical data quite well, as can be
seen by the examples given in fig. \ref{fig:g_local_bond_fitted}. The linear coefficients of the
quadratic fits correspond to critical temperatures of about \( 0.85 \) in the
\( 3d \) case. The ratio between the linear and the quadratic coefficients
is equivalent to an energy of about \( 0.6 \) (\( 2d \)) and \( 0.9 \)(\( 3d) \). The alternative
power fit results in an exponent \( \delta \sim 0.85 \) for both distributions. 

\begin{figure}
{\centering \epsfig{file=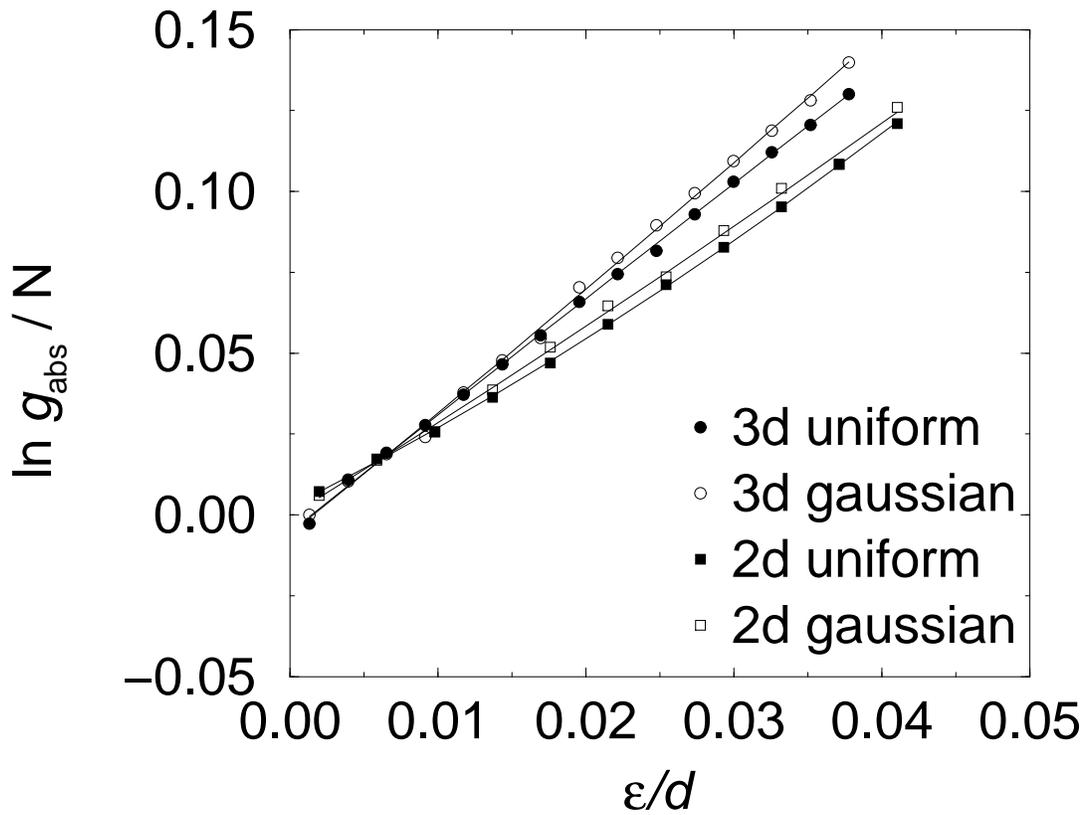, width=1\textwidth} \par}

\caption{Absolutely measured LDOS \protect\( g_{\mathrm{abs}}\protect \) versus energy per bond \protect\( \varepsilon /d\protect \) for \protect\( 2d\protect \) (squares)
and \protect\( 3d\protect \) (circles) systems with uniform (full symbols) or gaussian (open
symbols) distribution of interactions averaged over different realizations
of interactions and different valleys. The lines correspond to quadratic
fits.\label{fig:g_locabs_bond}}
\end{figure}

For the absolutely measured DOS it is not possible to map the results
for \( 2d \) to the results in \( 3d \) by taking into account the different coordination
numbers (Fig. \ref{fig:g_locabs_bond}). The planar systems result in a lower ALDOS compared
to the cubic systems. Moreover the power fits lead to exponents \( \delta  \),
which are very close to unity. The only exception is the \( 2d \) uniform
distributed system with \( \delta \sim 1.17 \). The quadratic fits result in ratios between
the linear and the quadratic coefficients larger than \( 1.0 \) per spin (\( 2d \)
gaussian) and larger than \( 2.0 \) per spin (\( 3d \)), which is almost the inverse
ground-state energy per spin. The exception is again the \( 2d \) uniform
distributed system with a ratio of about \( 0.3 \) per spin. All in all the
ALDOS grows almost exponentially with energy and the linear coefficients
correspond to temperatures of about \( 0.82 \) in \( 3d \) and \( 0.71 \) or \( 0.87 \) for \( 2d \) systems
with gaussian or uniform distribution, respectively.

\begin{figure}
{\centering \epsfig{file=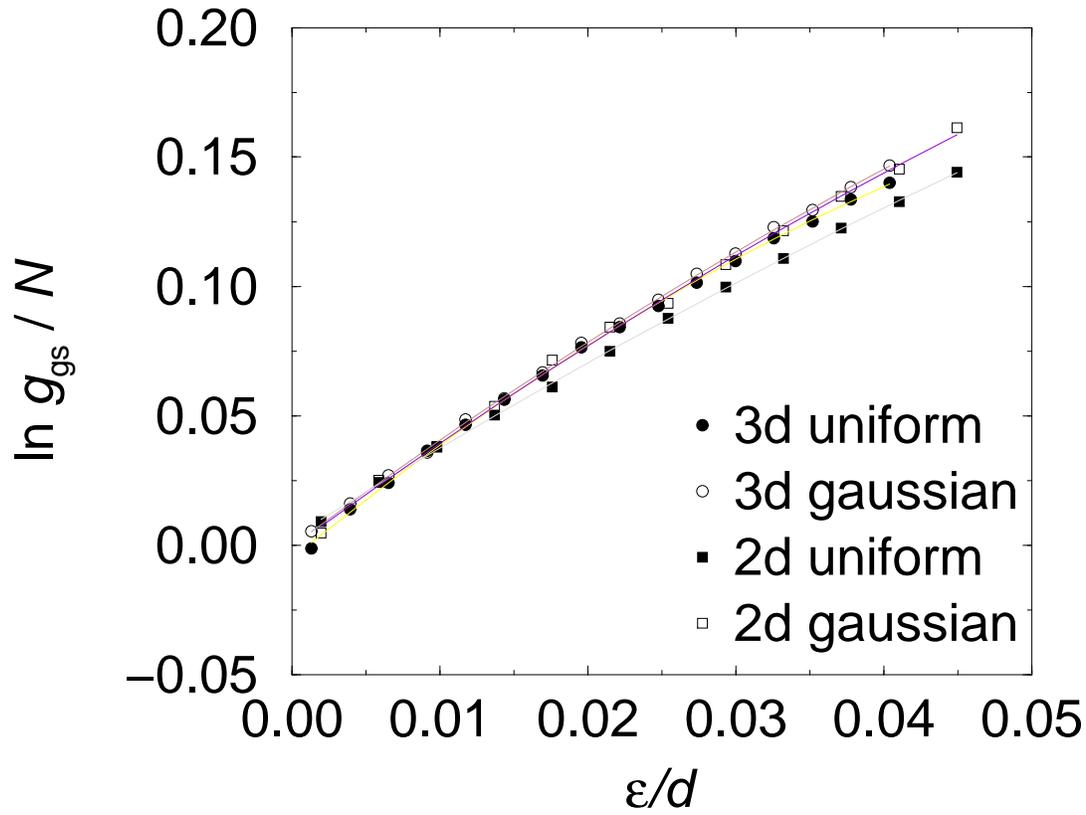, width=1\textwidth} \par}

\caption{Local density of states in the ground-state valleys \protect\( g_{\mathrm{gs}}\protect \) versus energy
per bond \protect\( \varepsilon /d\protect \) for \protect\( 2d\protect \) (squares) and \protect\( 3d\protect \) (circles) systems with uniform (full
symbols) or gaussian (open symbols) distribution of interactions averaged
over different realizations of interactions and different valleys.
The lines correspond to quadratic fits.\label{fig:g_locgs_bond}}
\end{figure}

For the averaged LDOS of the ground-state valleys the \( 2d \) uniform distributed
case seems to be an exception, too (Fig. \ref{fig:g_locgs_bond}). It is not clear, whether
this is really an effect or just a problem of the statistical errors.
The \( g_{\mathrm{gs}} \) versus energy per bond curves for the other cases agree quite
well. The power fits lead to an exponent \( \delta  \) between \( 0.77 \) and \( 0.89 \). According
to the ratio of the linear and the quadratic term of the quadratic
fits the deviations from the linear behaviour are of the order unity
for energies between \( 0.6 \) and \( 0.9 \) per spin. The linear terms correspond
to temperatures of about \( 0.5 \) (\( 2d \)) and \( 0.65 \) (\( 3d \)), which agree with the GDOS
values.

In fig. \ref{fig:g_compare_3du8} the different DOS measures are compared for the \( 3d \) uniform
distributed case. With increasing energy all valleys are joined successively
with more low-lying valleys. If there is essentially only the ground-state
valley left, the different DOS measures become equivalent. This seems
to be the case at an energy of about \( 0.13 \) per spin. For all lower energies
the global DOS, which counts the states in all existing valleys, is
certainly larger than the DOS of the ground-state valleys. Because
\( g_{\mathrm{abs}} \) is averaged over the ground-state valley and more high-lying valleys
and \( g_{\mathrm{abs}} \) is always lower than the ground-state valley DOS \( g_{gs} \), for absolutely
measured energies the local DOS of the high-lying valleys is smaller
than the GLDOS. On the other hand the relatively measured RLDOS \bfseries \( g_{\mathrm{rel}} \)\mdseries ,
which measures the DOS relatively to the minimal energy of a valley,
is always larger than the GLDOS. Therefore the more high lying valleys
must have larger local densities of states than the ground-state valley
measured relatively to the minimal energy of these valleys.

\section{Summary}

We investigated the global and the local DOS for square and cubic
Ising spin glass systems with a gaussian and with a uniform probability
distribution of interactions, respectively. The quantitative differences
between the \( 2d \) and the \( 3d \) systems are mostly caused by their different
coordination numbers. Although the first two moments of the chosen
distributions of interactions were set equal, the DOS for the gaussian
systems is slightly higher than for the uniform distributed systems.
However there is no significant qualitative difference. Therefore
it should be possible to use both distributions alternatively for
investigations of the state-space structure.

From the direct comparison of the different DOS measures it follows,
that at a given absolute energy most of the valleys have a lower LDOS
than the ground-state valley. On the other hand, the more high lying
valleys have a higher LDOS measured relatively to the minimal energy
of the considered valley. Thus we get a picture of the state space
with small energetically low-lying valleys which have high energy
barriers and wide energetically high-lying valleys with low energy
barriers.

\begin{figure}
{\centering \epsfig{file=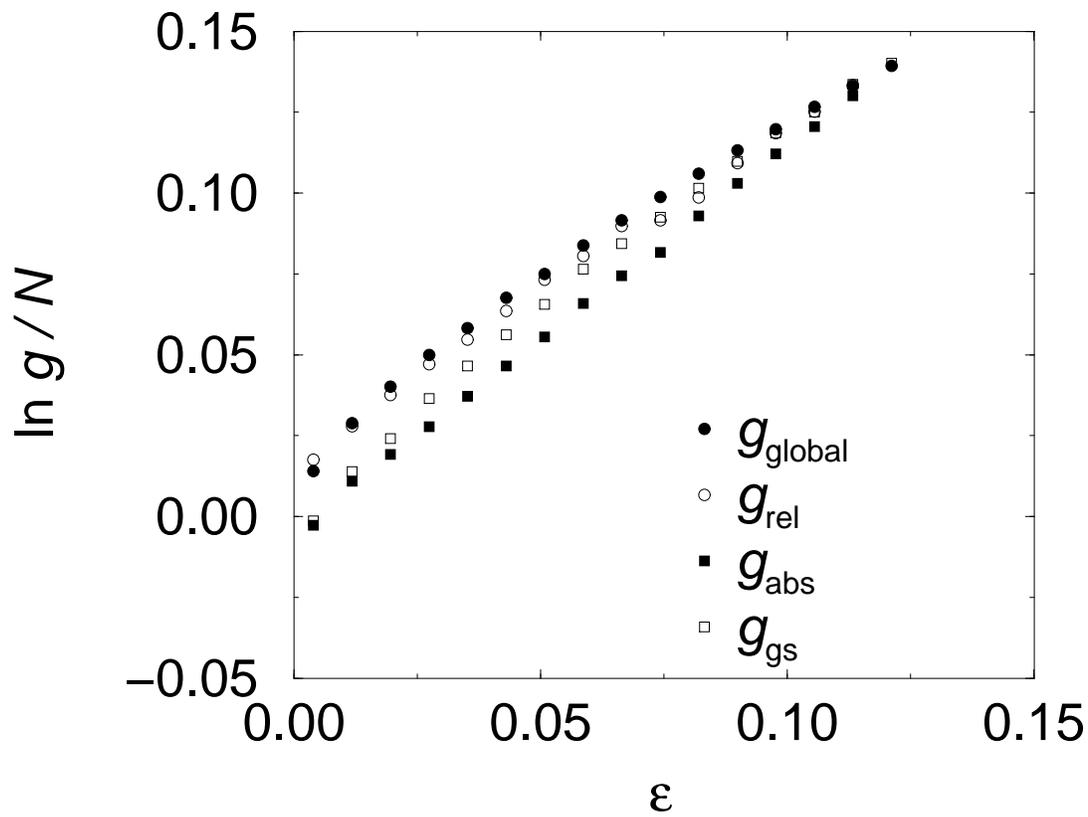, width=1\textwidth} \par}

\caption{Comparison of the different DOS measures for the \protect\( 3d\protect \) uniform distributed
case.\label{fig:g_compare_3du8}}
\end{figure}

The existence of a large ground-state valley in the system could explain,
why simple heuristic and approximative optimizing algorithms often
are able to find very good sub-optimal states in problems of this
kind. In a first approximation the probability to find the ground-state
valley of a system by a random search at a given energy is defined
by the ratio between the DOS of the ground-state valley and the global
DOS. In fig. \ref{fig:g_compare_3du8} this ratio is for high energies close to unity and
decreases for lower energies. Therefore a simple search algorithm
can easily find the true ground-state valley at high energies (or
high temperatures). With decreasing energy or temperature the chance
to hit the right sub-valley decreases. Thus the algorithm will find
sub-optimal solutions, but not necessarily the optimal state. 

A second feature of the state space picture seen here is, that for
valleys which start at a high energy the RLDOS grows faster with energy
than for valleys which start at a lower energy. As the LDOS should
determine most of the non-equlibrium thermodynamic properties seen
in real or computer experiments, these properties will depend on the
energy range at which the system is investigated. This should be kept
in mind while approximating ground-state or low-temperature properties
by the investigation of energetically high-lying valleys.

The more detailed quantitative analysis of the DOS shows that only
the absolutely measured local DOS \( g_{\mathrm{abs}} \) grows almost exponentially. All
other measures for the local DOS and the global DOS grow clearly subexponentially.
In all these cases the applied two trial fits with a quadratic and
a power ansatz, respectively, describe the numerical data for the
logarithm of the DOS quite well.

For the quadratic fits the corrections to the linear behaviour become
of the order unity for energies of about \( 0.5 \) per spin. The linear coefficients
correspond to temperatures, which are for the \( 3d \) systems in the region
of the transition temperature found for the gaussian systems \cite{T_c_gaussian}.
However, although there should only be a zero-temperature transition
in \( 2d \), the temperatures resulting from the quadratic fits are about
\( 0.5 \). The power fits differ in the absolute terms and the coefficients
of the power terms. The exponents are with values about \( 0.7 \) significantly
different to a linear behaviour. Therefore there is at least no sharply
defined transition temperature, below which a system is trapped in
a valley. 

The question whether the form of the local DOS leads to a transition
at all remains unclear. To decide this, a further analysis of the
occupation probability of a valley with respect to the distribution
of energy barriers of this valley would be necessary. Furthermore
the barrier distribution combined with the density of local minima
will give a better understanding of the physical meaning and the connections
between the different LDOS measures.

It should be noted here that the definition of the different DOS measures
does not depend on the underlying model. Therefore the quantitative
analysis of these properties should give a better insight in the state
space structure of similar physical and optimization problems, too.

\begin{lyxbibliography}

\bibitem{Aging1}J. Hammann, M. Lederman, M. Ocio, R. Orbach and E. Vincent,
Physica A \bfseries 185 \mdseries (1992) 278.

\bibitem{Aging2}J.-O. Andersson, J. Mattson and P. Nordblad, Phys. Rev. B
\bfseries 48 \mdseries (1993) 13977.

\bibitem{SK}D. Sherrington and S. Kirckpatrick, Phys. Rev. Lett. \bfseries 32
\mdseries (1975) 1792.

\bibitem{Nemoto}K. Nemoto, J. Phys. A \bfseries 21 \mdseries (1988)
L287.

\bibitem{Vertechi}D. Vertechi and M. A. Virasoro, J. Phys. \bfseries 50
\mdseries (1989) 2325.

\bibitem{Ultrametricity}M. Mezard, G. Parisi, N. Sourlas, G. Toulouse and
M. A. Virasoro, Phys. Rev. Lett. \bfseries 52 \mdseries (1984)
1156.

\bibitem{Klotz}T. Klotz and S. Kobe, Acta Phys. Slovaca \bfseries 44 \mdseries (1994)
347.

\bibitem{Sibani}P. Sibani and P. Schriver, Phys. Rev. B \bfseries 49 \mdseries (1993)
6667; P. Sibani and J. O. Andersson, Physica A \bfseries 206 \mdseries (1994)
1.

\bibitem{sven}J. C. Sch\"on , J. Phys. A \bfseries 30 \mdseries (1997)
2367.

\bibitem{BB}A. Hartwig, F. Daske and S. Kobe, Comp. Phys. Commun. \bfseries 32
\mdseries (1984) 133.

\bibitem{T_c_gaussian}R. N. Bhatt and A. P. Young, Phys. Rev. B \bfseries 37
\mdseries (1988) 5606.

\end{lyxbibliography}

\newpage
\listoffigures

\end{document}